\documentclass{pasj00}
\SetRunningHead{T. Fujinaga et al.}{Suzaku observation of HESS J1702--420}

\usepackage{times}

\begin{document}

\title{Suzaku observation of the unidentified VHE gamma-ray source HESS J1702--420}
\author{Takahisa~Fujinaga,\altaffilmark{1,2} Aya~Bamba,\altaffilmark{3} Tadayasu~Dotani,\altaffilmark{1,2} Masanobu~Ozaki,\altaffilmark{1} \\ Gerd~P\"uhlhofer,\altaffilmark{4} Stefan~Wagner,\altaffilmark{5} Olaf~Reimer,\altaffilmark{6,7} Stefan~Funk,\altaffilmark{7} and Jim~Hinton\altaffilmark{8}}
\altaffiltext{1} {Institute of Space and Astronautical Science/JAXA, 3-1-1 Yoshinodai, Chuo-ku, Sagamihara, Kanagawa 252-5210}
\email{fujinaga@astro.isas.jaxa.jp}
\altaffiltext{2}{Department of Physics, Tokyo Institute of Technology, 2-12-1 Ookayama, Meguro-ku, Tokyo 152-8550}
\altaffiltext{3}{Department of Physics and Mathematics, Aoyama Gakuin University, 5-10-1 Fuchinobe, Chuo-ku, Sagamihara, Kanagawa 252-5258}
\altaffiltext{4}{Institut f\"ur Astronomie und Astrophysik, Universit\"at T\"ubingen, Sand 1, D 72076 T\"ubingen, Germany}
\altaffiltext{5}{Landessternwarte, Universit\"at Heidelberg, K\"onigstuhl, D 69117 Heidelberg, Germany}
\altaffiltext{6}{Institut f\"ur Astro- und Teilchenphysik, Leopold-Franzens-Universit\"at Innsbruck, A-6020 Innsbruck, Austria}
\altaffiltext{7}{SLAC National Accelerator Laboratory, 2575 Sand Hill Road, Menlo Park, CA 94025, USA}
\altaffiltext{8}{University of Leicester, Leicester LE1 7RH, UK} 

\maketitle

\begin{abstract}
A deep X-ray observation of the unidentified very high energy (VHE) gamma-ray source HESS J1702--420, for the first time, was carried out by Suzaku. 
No bright sources were detected in the XIS field of view (FOV) except for two faint point-like sources. 
The two sources, however, are considered not to be related to HESS J1702--420, because their fluxes in the 2--10~keV band ($\sim 10^{-14}~{\rm erg}~{\rm s}^{-1}~{\rm cm}^{-2}$) are $\sim 3$ orders of magnitude smaller than the VHE gamma-ray flux in the 1--10~TeV band ($F_{\rm TeV} = 3.1~\times~10^{-11}~{\rm erg~s}^{-1}~{\rm cm}^{-2}$). 
We compared the energy spectrum of diffuse emission, extracted from the entire XIS FOV with those from nearby observations. 
If we consider the systematic error of background subtraction, no significant diffuse emission was detected with an upper limit of  $F_{\rm X} < 2.7~\times~10^{-12}~{\rm erg~s}^{-1}~{\rm cm}^{-2}$ in the 2--10~keV band for an assumed power-law spectrum of $\Gamma=2.1$ and a source size same as that in the VHE band. 
The upper limit of the X-ray flux is twelve times as small as the VHE gamma-ray flux. 
The large flux ratio ($F_{\rm TeV}/F_{\rm X}$) indicates that HESS J1702--420 is another example of a ``dark'' particle accelerator. 
If we use a simple one-zone leptonic model, in which VHE gamma-rays are produced through inverse Compton scattering of the cosmic microwave background and interstellar far-infrared emission, and the X-rays via the synchrotron mechanism, an upper limit of the magnetic field ($1.7~\mu{\rm G}$) is obtained from the flux ratio. 
Because the magnetic field is weaker than the typical value in the Galactic plane ($3-10~\mu{\rm G}$), the simple one-zone model may not work for HESS J1702--420 and a significant fraction of the VHE gamma-rays may originate from protons.
\end{abstract}

\section{Introduction}
\label{sec:intro}

The origin of cosmic rays is an unsolved problem since their discovery in the early 20th century. 
The most widely accepted mechanism for the cosmic ray acceleration up to $10^{15.5}$~eV (the knee energy) is diffusive shock acceleration at the forward shocks of supernova remnants (SNRs) (e.g., \cite{bell1978}; \cite{blandford1978}). 
Young SNRs sometimes show synchrotron X-ray emission, which can be the direct evidence of accelerated electrons exceeding $\sim 1$~TeV (e.g., \cite{koyama1995}; \cite{koyama1997}; \cite{bamba2000}; \cite{bamba2001}; \cite{slane2001}). 
However, no clear evidence of proton acceleration was found through the X-ray observations. 

Recently, a very high energy (VHE) gamma-ray telescope, High Energy Stereoscopic System (H.E.S.S.), was used to carry out survey observations along the Galactic plane (\cite{aharonian2005a}; \cite{aharonian2006}). 
The surveys led to the discovery of many new VHE gamma-ray sources. 
Because of their distribution along the Galactic plane, they are considered to be located in the Galaxy. 
In fact, two thirds of the sources were found to be associated with SNRs or pulsar wind nebulae (PWNe). 
For most of the remaining sources, no counterparts have yet been found in other wavelengths, and their nature is still unclear. 
Because VHE gamma-rays are produced either through inverse Compton scattering of low-energy photons by energetic electrons, or through the decay of pions produced by collisions of high energy protons with interstellar medium, high energy particles are surely involved in the VHE gamma-ray sources. 
They have therefore been referred to as 
``dark'' particle accelerators (\cite{aharonian2005a}; \cite{ubertini2005}, for example). 
If high energy electrons are present, they may be traced through the observations of synchrotron X-ray emission. 

HESS J1702--420 is one of the brightest sources among the unidentified (unID) VHE gamma-ray sources (\cite{aharonian2006}; \cite{aharonian2008}). 
Its spatial extent is asymmetric ranging from \timeform{15'} through \timeform{30'}. 
Its flux in the 1 -- 10~TeV band is  $3.1~\times~10^{-11}~{\rm erg~s}^{-1}~{\rm cm}^{-2}$ with a characteristic power-law spectrum with $\Gamma =2.07 \pm 0.08$. 
An SNR G344.7--0.1 and a pulsar PSR J1702--4128 are located in the outskirts of this source. 
\citet{aharonian2008} concluded that G344.7--0.1 is not associated to HESS J1702--420 because the angular size of the SNR is too small and the estimated distance of SNR (14~kpc, \cite{dubner1993}) is very large. 
On the other hand, PSR J1702--4128 may be related to HESS J1702--420 (\cite{gallant2007}). 
Also, a hint of extended X-ray emission characteristic of PWNe was obtained by Chandra (\cite{chang2008}). 
However, given the large angular offset of $35'$ between the pulsar and the peak of the VHE gamma-ray source, which corresponds to $\simeq 50$~pc for the distance of 5.1~kpc to the pulsar (\cite{guseinov2004}) , and the 3 orders of magnitude difference between the X-ray PWN energy flux ($2~\times~10^{-14}~{\rm erg~s}^{-1}~{\rm cm}^{-2}$ in the 0.3 -- 10~keV band) and the VHE gamma-ray flux, we consider the association of this source rather weak. 

A deep observation of the sky field including HESS J1702--420 has not been carried out so far, and only survey results are available. 
No plausible counterpart is found in the Galactic SNR catalog (\cite{green2009}). 
1FGL J1702.4--4147c, located 14~arcmin away from HESS J1702--420, is listed on the Fermi first source catalog (\cite{abdo2010}). 
They reported that the diffuse background model needs to be improved on the Galactic plane and the position of the GeV source depends on this model. 
Thus, it is unclear that 1FGL J1702.4--4147c is a counterpart of HESS J1702--420. 
In order to search for an X-ray counterpart, we made a deep X-ray observation of HESS J1702--420 with Suzaku. 
In this paper, we report the results of the X-ray observation. 

\begin{figure}
  \begin{center}
    \FigureFile(75mm,56mm){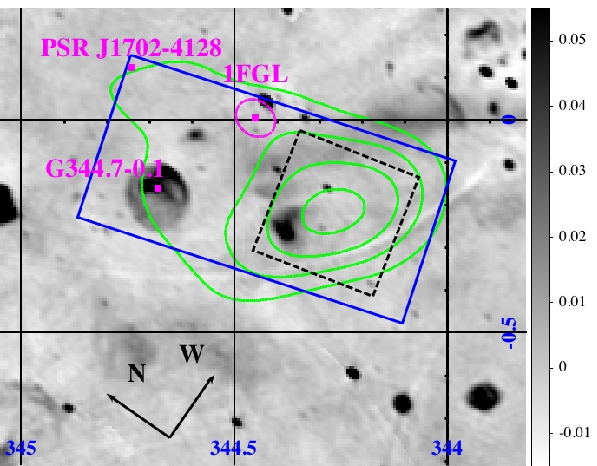}
  \end{center}
  \caption{  
  The 843~MHz Molongo radio image of the HESS J1702--420 field (\cite{green1999}) is shown in gray scale in units of Jy beam$^{-1}$, in Galactic coordinates.
  The green contours show the H.E.S.S. intensity map (\cite{aharonian2006}) in linear scale.   
  The $1\sigma$ error region of the Fermi source, 1FGL J1702.4--4147c, is indicated by a magenta circle with the label ``1FGL'' (\cite{abdo2010}).
  The black dashed box represents the FOV of Suzaku XIS. The emission region of VHE gamma-rays is approximated by the blue solid rectangle of $0\fdg8 \times 0\fdg4$, when we constrain the total X-ray emission from the source.}
  \label{fig:hessimage}
\end{figure}

\section{Observation}
\label{sec:obs}

We performed a deep X-ray observation of the sky field including HESS J1702--420 with Suzaku (\cite{mitsuda2007}) from March 25 through 30, 2008. 
The field of view (FOV) was centered at $(l,b)=(344\fdg26,-0\fdg22)$ as shown in Figure \ref{fig:hessimage} as the dashed black box. 
The observation log is listed in Table \ref{tab:obslog}. 

Suzaku is equipped with two types of detectors: four sets of X-ray imaging spectrometers (XIS0--XIS3: \cite{koyama2007}) and a hard X-ray detector (HXD: \cite{takahashi2007}; \cite{kokubun2007}). 
XIS1 uses a back-illuminated (BI) CCD while the others front-illuminated (FI) CCDs. 
Among the four sensors, XIS2 was not usable during the observation. 
XIS was operated in the normal mode without any options. 
The spaced-row charge injection (SCI: \cite{nakajima2008}) was used to reduce the effects of radiation damage. 

\begin{longtable}{*{7}{c}}
\caption{Log of the source and background observations}
\label{tab:obslog}
\hline
\hline
 & Obs ID & Observation date & Aim point\footnotemark[$*$] & Exposure & FI CCD & SCI \\
 & & & ($l, b$) & (ks) &  Count rate\footnotemark[$\dagger$] & \\
\hline
\multicolumn{1}{l}{HESS J1702--420} & 502049010  & 2008/03/25--2008/03/30 & ($344\fdg26, -0\fdg220$) & 216 & $1.62 \pm 0.01$ & On\\
\hline 
\multicolumn{1}{l}{Background 1} & 100028010 & 2005/09/19--2005/09/20 & ($332\fdg40, -0\fdg150$) & 45 & $1.33 \pm 0.03$ & Off\\
\multicolumn{1}{l}{Background 2} & 100028020 & 2005/09/18--2005/09/19 & ($332\fdg00, -0\fdg150$) & 21 & $1.39 \pm 0.04$ & Off\\
\hline
\multicolumn{7}{l}{\footnotemark[$*$] In Galactic coordinates.} \\
\multicolumn{7}{l}{\footnotemark[$\dagger$] In units of $10^{-1}$~counts~s$^{-1}$ in the 4--8 keV band, with 90\% statistical uncertainty.} 
\endlastfoot
\end{longtable}

\section{Analysis and Results}
\label{sec:ana}

We concentrate on the analysis of XIS, because the HXD data are largely contaminated by the Galactic ridge X-ray emission (GRXE: cf., \cite{koyama1989}; \cite{yuasa2008}; \cite{yamauchi2009}), which are difficult to subtract correctly due to the location of HESS J1702--420 near the Galactic plane. 
It was clear from the quick-look analysis of XIS that no bright source was present in the FOV. 
Therefore, accurate background subtraction is crucial to identify a possible counterpart. 
For this purpose, we selected two sets of archive data which are useful to estimate the background of this observation. 
Considering the postion dependence of the GRXE, the data were selected to satisfy the following criteria: (1)~$300^\circ~\leq~l~\leq~350^\circ, -0\fdg25~\leq~b~\leq~-0\fdg15$, (2) free from bright sources, and (3) observed with normal clocking mode. 
Hereafter, we refer to these two sets of data as background~1 and background~2, respectively.  
Their observation logs are also listed Table \ref{tab:obslog}. 
All the four XIS were operational during in the background observations. 

We used version 2.2.7.18 of the processed data for HESS J1702--420 and version 2.0.6.13 for the background data, respectively. 
The differences between these versions on XIS are the calibration data for SCI and that for burst options, which are negligible for the analyses in this paper. 
The data were analyzed with the HEADAS software version 6.7 and XSPEC version 12.5.1. 
We used the cleaned event file created by the Suzaku team. 
The resultant effective exposure is listed in Table~\ref{tab:obslog}. 

\subsection{Images}
\label{subsec:image}

\begin{figure*}
  \begin{center}
    \FigureFile(80mm,60mm){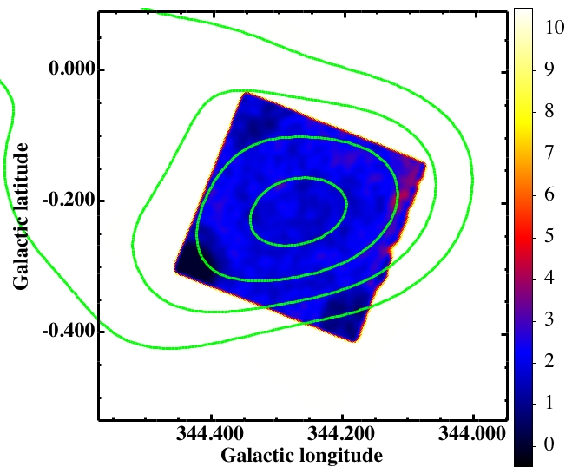} \ \ \ \ \
    \FigureFile(80mm,60mm){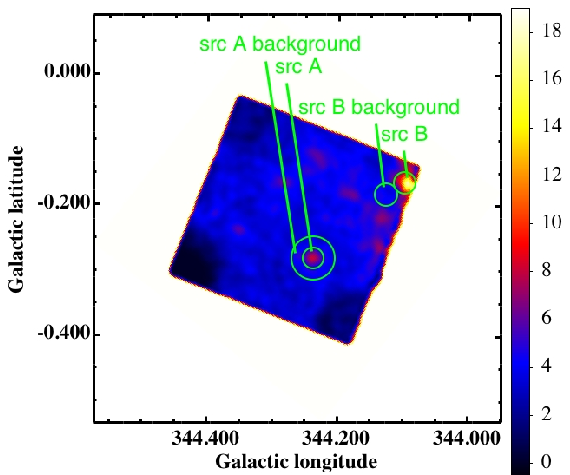}
  \end{center}
  \caption{Suzaku XIS images of HESS J1702--420 field in the 0.5--2~keV band (left) and the 2--8~keV band (right).  
  Definition of the color scale is indicated in the right-hand-side of each panel in unit of counts/pixel. 
  These images were binned to a pixel size of \timeform{8''}, and were smoothed with a Gaussian of $\sigma=$~\timeform{1'.0}. 
  The vignetting is corrected after the NXB subtraction. In the left image, the VHE gamma-ray intensity map is overlaid in the green contour (the same as Figure \ref{fig:hessimage}). The regions described in the right image are used for the point source analysis (see Table \ref{tab:srcAsrcB}).}
  \label{fig:xisimage}
\end{figure*}

We first calculated a vignetting corrected image. 
For this purpose, we generated an image of the non X-ray background (NXB) using the FTOOL {\tt xisnxbgen} (\cite{tawa2008}). 
The NXB of XIS generally depends on the cut-off rigidity (COR). 
Thus {\tt xisnxbgen} utilize the NXB database sorted by COR. 
It calculates the COR distribution of the data and generates a background for the same distribution of COR. 
After subtracting the NXB image, we corrected vignetting using the vignetting map created by the FTOOL {\tt xissim} (\cite{ishisaki2007}). 
We then summed up all the images of XIS0, XIS1 and XIS3. 
The results are shown in Figure \ref{fig:xisimage} for both the soft and hard energy bands. 

No clear point source is seen in the soft energy band, whereas two significant sources are found in the hard energy band. 
Hereafter, the two sources are referred to as src A and src B. 
They are consistent with point sources if we consider the fluctuation of the pointing direction of Suzaku, which is at most $\simeq$~\timeform{1'} (\cite{serlemitsos2007}; \cite{uchiyama2008}), and a typical half power diameter of the point spread function ($\simeq$~\timeform{2'}, \cite{serlemitsos2007}). 
Table \ref{tab:srcAsrcB} summarizes the celestial position, total counts from the source and the statistical significance of the two point sources. 
The total counts were calculated for the source region defined by a circle with a radius of \timeform{1'}, and the background region next to each source region (see Figure \ref{fig:xisimage}). 
The statistical significances of src A and src B were $8\sigma$ and $5\sigma$, respectively. 
According to the SIMBAD database, no catalogued source exists within \timeform{2'} from these sources. 

\begin{figure}
  \begin{center}
    \FigureFile(80mm,50mm){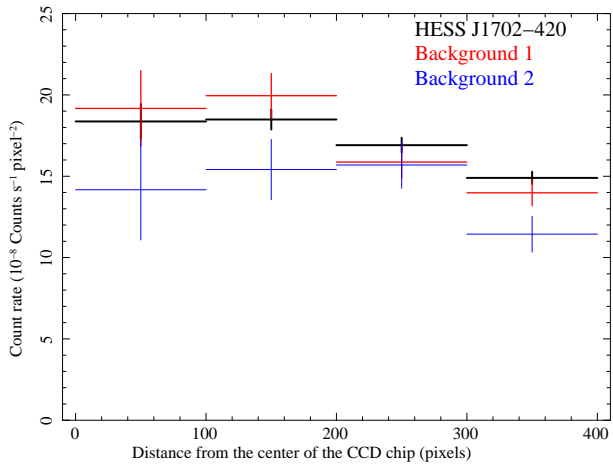}
  \end{center}
  \caption{The 4--8~keV radial profile of FI CCD data in detector coordinates with 90\% error bars. NXBs were subtracted.}
  \label{fig:radial}
\end{figure}

We also looked for a diffuse emission in excess of the background. 
The NXB subtracted radial profile are shown in Figure~\ref{fig:radial}. 
All the three data have similar profile each other, which 
means that no bright diffuse emission is present in the XIS FOV. 
More careful analysis is needed to quantify the faint diffuse emission, which is given in section \ref{subsec:diffuse}. 

\begin{table}
  \caption{Summary of the src A and src B}
  \label{tab:srcAsrcB}
  \begin{center}
    \begin{tabular}{lcc}
    \hline 
    \hline 
     & src A & src B \\
    \hline
    $l$ (degree) & $344\fdg237$ & $344\fdg093$ \\
    $b$ (degree) & $-0\fdg281$ & $-0\fdg167$ \\
    \hline
    The total counts of FI CCDs & & \\
    \ \ \ \ The source region\footnotemark[$*$] & $1692$ & $1236$ \\
    \ \ \ \ The background region\footnotemark[$*$] & $1269$ & $1008$ \\
    \hline 
    Excess \footnotemark[$*$] & 423 & 228  \\
    Significance & 8$\sigma$ & 5$\sigma$ \\
    \hline
    \multicolumn{3}{l}{Note: The energy band is in 0.5--10~keV.} \\
    \multicolumn{3}{l}{\ \ \ \ \ \ \ \ \ \ The effective exposure time is 216~ks.} \\
    \multicolumn{3}{l}{\footnotemark[$*$] In units of counts per a circle with radius of \timeform{1'}.} \\
    \end{tabular}
  \end{center}
\end{table}

\subsection{Spectra of point sources}
\label{subsec: point}

Since the poor statistics prevented us from detailed spectral analysis of the two point sources, we converted their count rates into the X-ray flux using {\tt WebPIMMS}\footnote{http://heasarc.gsfc.nasa.gov/Tools/w3pimms.html}. 
Because the mirror vignetting (0.8 for src A and 0.7 for src B) is not considered in {\tt WebPIMMS}, it was corrected separately. 
We assumed that they have the same intrinsic spectrum of a power-law with a photon index of 2.1 (same as HESS J1702--420) modified by the interstellar or circumstellar absorption of $N_{\rm H}~=~1.5~\times~10^{22}~{\rm cm}^{-2}$ (\cite{kalberla2005}). 
The X-ray fluxes in the 2--10~keV band were $(3.0 \pm 0.6)~\times~10^{-14}~{\rm erg~s}^{-1}~{\rm cm}^{-2}$ (src A) and $(1.9 \pm 0.7)~\times~10^{-14}~{\rm erg~s}^{-1}~{\rm cm}^{-2}$ (src B), where the errors are in 90\% confidence limit. 

\subsection{Diffuse emission}
\label{subsec:diffuse}

In this subsection, we look for an excess diffuse emission, which may be associated to HESS J1702--420. 
We assume for simplicity that the diffuse emission is uniformly extended in the XIS FOV. 
For this purpose, we calculate the energy spectrum (including background) of the entire XIS FOV and compare it with that of the background observations. 
Below 4~keV, Background 1 was contaminated by an SNR, RCW 103. 
Thus we analyzed only the data in the 4--8~keV band. 
It is noteworthy that the observation of HESS J1702--420 was carried out with SCI-on, whereas the background observations with SCI-off. Because dead areas are introduced in the XIS image with SCI-on, they need to be incorporated in the analysis using appropriate response files. 
Two kinds of response files, i.e. redistribution matrix files (RMFs) and ancillary response files (ARFs), were made for each spectrum using the FTOOL {\tt xisrmfgen} and {\tt xissimarfgen} (\cite{ishisaki2007}), respectively. 
The dead area is included in the ARFs. 

The background data of XIS consist of 3 components, i.e. NXB, cosmic X-ray background (CXB), and GRXE. 
When we compare the background data of different observations, it is important to note the time or spatial variations of these background components. 
NXBs were generated using {\tt xisnxbgen} as explained in section \ref{subsec:image}. 
The FTOOL, {\tt xisnxbgen}, reproduces the long-term variation of NXB by taking average of data spanning $\pm~150$~days centered on the observation date of the input data. 
This is true for the observation of HESS J1702--420. 
However, this was difficult for the background data because they were obtained just after the launch of Suzaku.  
Thus the data used to take an average span only 30~days before the observation (and 150~days after the observation). 
However, the difference of the averaging span does not affect the reproducibility of the NXB, because the NXB is known to be stable over a long period of time (\cite{tawa2008}).
Based on \citet{tawa2008}, we estimate the systematic error of NXB is 5.9~\% in 4--8~keV (at 90~\% confidence limit). 
The NXB subtracted energy spectra for background~1 and 2 were summed up (background~1+2) to use the later analysis. 
The GRXE is constant in time, but depends on the Galactic coordinate, mostly on latitude. 
HESS J1702--420 is separated from background 1+2 by 12 degree along to the Galactic longitude, while their Galactic latitudes are almost same (see Table~\ref{tab:obslog}). 
According to \citet{revnivtsev2006}, angular separation of 12 deg along the Galactic longitude in the vicinity of HESS J1702--420 may cause systematic change of GRXE by $\simeq 20$~\%. 
In addition to this, GRXE has small scale fluctuation. 
Thus, a careful analysis may be needed. 
In what follows, we compare the GRXE between HESS J1702--420 and background 1+2. 
Then, we proceed to the search for the excess emission from HESS J1702--420. 

\begin{figure*}
  \begin{center}
    \FigureFile(80mm,50mm){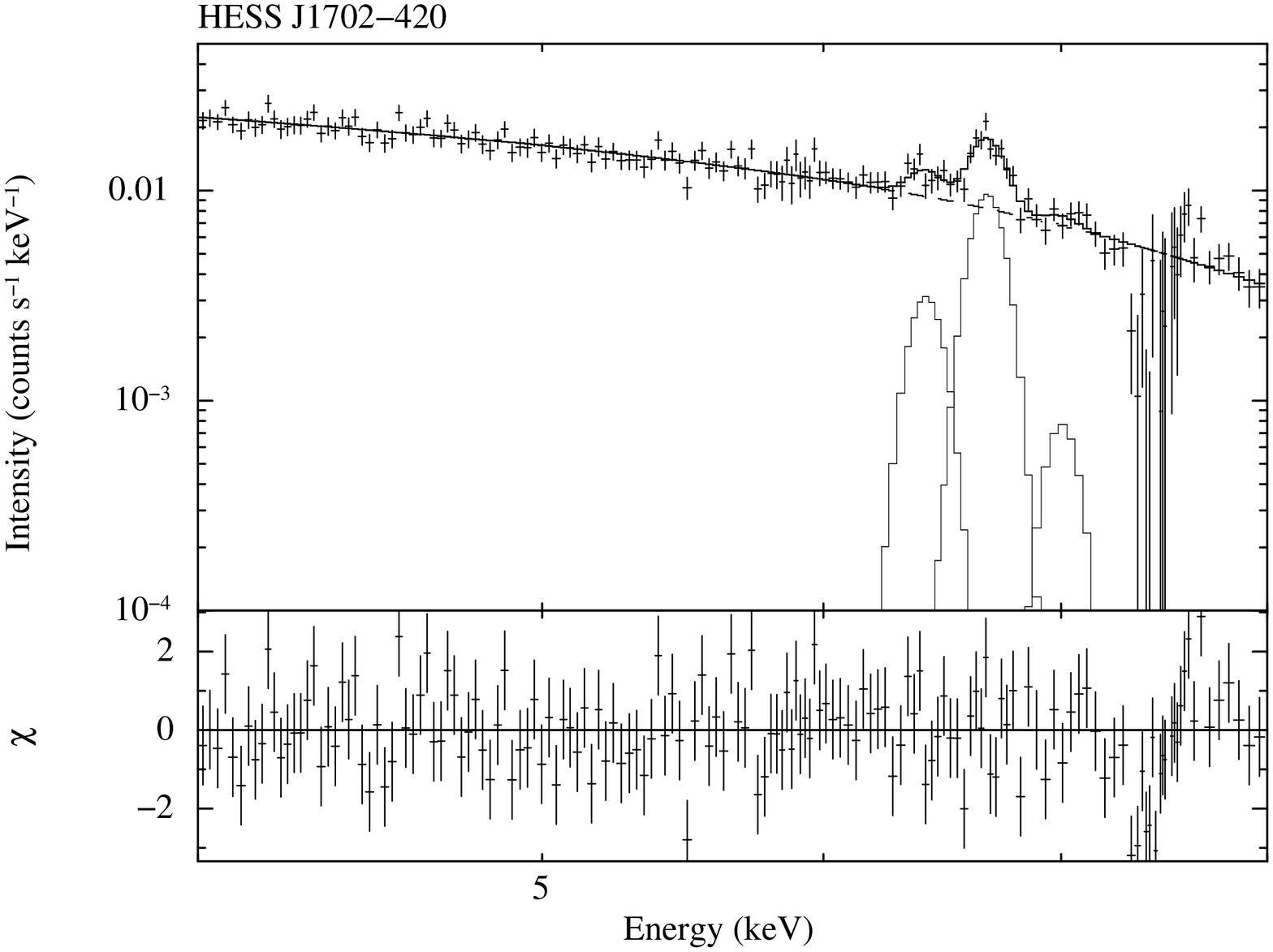}
    \ \ \ \ \ 
    \FigureFile(80mm,50mm){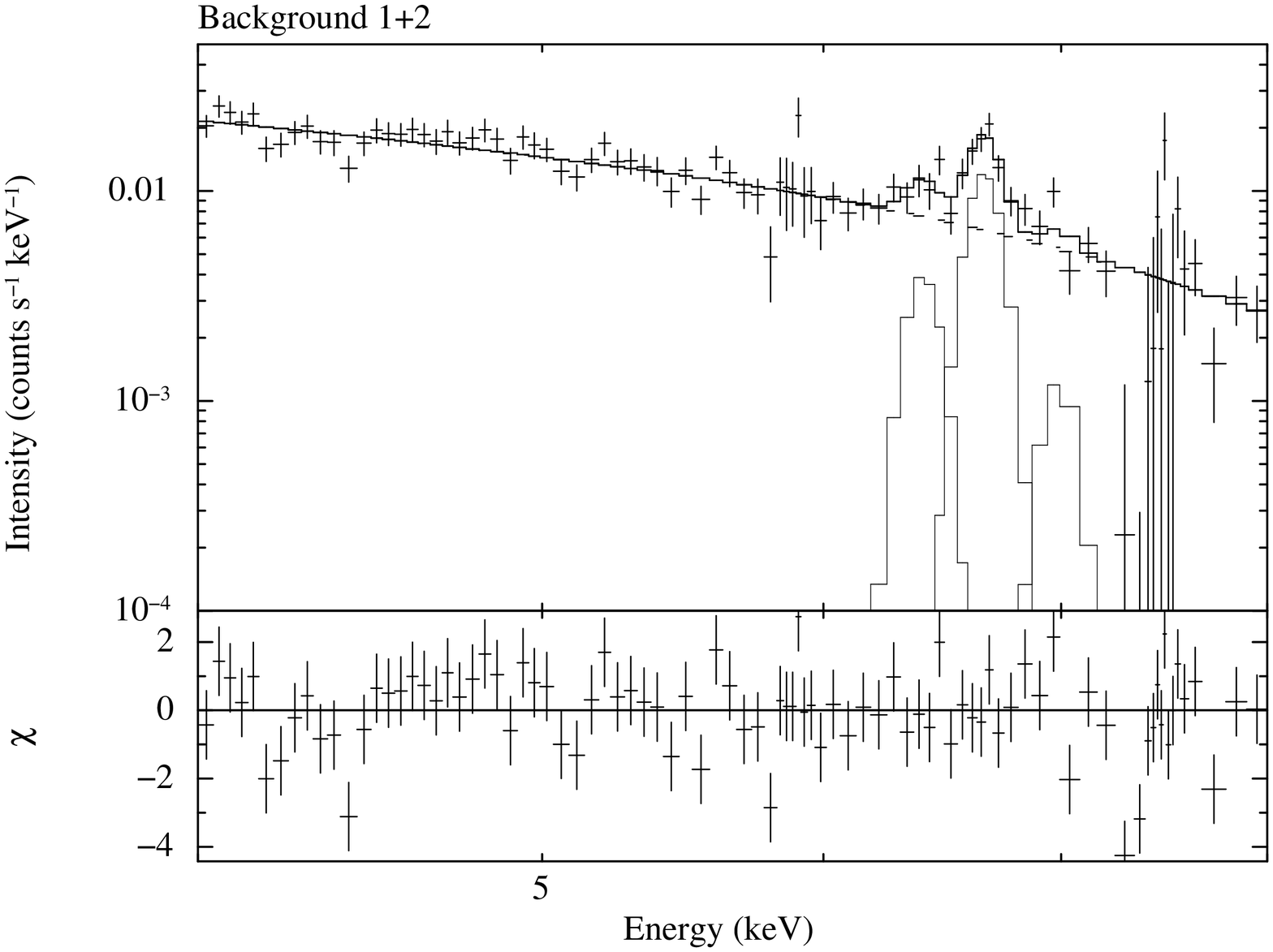}
  \end{center}
  \caption{XIS FI spectra of HESS J1702--420 (left) and Background 1+2 (right) in the 4--8~keV band. The dashed line and solid lines show the power-law component and the iron line features.}
  \label{fig:grxespec}
\end{figure*}

At first, we compared the NXB subtracted spectra of HESS J1702--420 with those of Background 1+2. 
We adopted a simple model spectrum, a power-law with three gaussians, which is appropriate to represent a sum of the CXB and GRXE. 
We fitted HESS J1702--420 and background spectra separately. 
The results are shown in Figure \ref{fig:grxespec} and the best-fit parameters are summarized in Table \ref{tab:grxest}. 
The flux of the power-law is not consistent with each other, which may be due to the systematics of the background subtraction and/or the presence of excess emission. 
The three gaussians are considered to be associated to the GRXE and their fluxes are a good measure of the GRXE flux (\cite{yamauchi2009}). 
As explained in the previous paragraph, the GRXE flux around HESS J1702--420 may be larger than that around background 1+2 by 20~\% due to its global variation. 
However, the gaussian fluxes do not follow this trend. 
In fact, they are consistent with each other within the statistical errors. 
We consider that this is due to the small scale fluctuation of the GRXE and the intrinsic difference of the GRXE fluxes may be smaller than 20~\%. 
In the present analysis, we assume conservatively that the systematic error of GRXE is at most 20~\%. 

\begin{longtable}{*{4}{c}}
\caption{The best-fit parameters of the HESS J1702--420 and the background spectra.}
\label{tab:grxest}
    \hline 
    \hline 
     & & HESS J1702--420 & Background 1+2 \\
    \hline
     \multicolumn{1}{l}{Power-law} & \multicolumn{1}{l}{$\Gamma$} & $1.6 \pm 0.1$ & $2.1 \pm 0.1$ \\
     & \multicolumn{1}{l}{Flux\footnotemark[$*$]} & $2.7 \pm 0.1$ & $2.3 \pm 0.1 $ \\
     \multicolumn{1}{l}{Neutral-Fe} & \multicolumn{1}{l}{Energy~(keV)} & $6.40 \pm 0.03$ & $6.46 ^{+0.03} _{-0.04}$ \\
     &  \multicolumn{1}{l}{Intensity\footnotemark[$\dagger$]} & $ 3.9 \pm 1.3 $ & $ 6.1 ^{+1.6}_{-1.7}$ \\
     \multicolumn{1}{l}{Fe$_{\rm XXV}$} &  \multicolumn{1}{l}{Energy~(keV)} & $ 6.68 \pm 0.01$ & $6.68 \pm 0.01$ \\
     &  \multicolumn{1}{l}{Intensity\footnotemark[$\dagger$]} & $ 12.7 \pm 1.5 $ & $ 13.5 ^{+1.9}_{-1.8}$ \\
     \multicolumn{1}{l}{Fe$_{\rm XXVI}$} &  \multicolumn{1}{l}{Energy~(keV)} & $6.97$~(fixed) & $6.97$~(fixed) \\
     &  \multicolumn{1}{l}{Intensity\footnotemark[$\dagger$]} & $1.0 ^{+2.0}_{-1.0}$ & $ 4.0 \pm 1.6$ \\
    \hline
    $\chi ^2/$d.o.f.  & & $193.37/163$ & $185.15/124$ \\
    \hline 
    \multicolumn{4}{l}{Notes. Errors are in 90\% confidence region.} \\
    \multicolumn{4}{l}{\footnotemark[$*$] In units of  $10^{-12}$~erg~s$^{-1}$~cm$^{-2}$ for 4--8~keV.}\\
     \multicolumn{4}{l}{\footnotemark[$\dagger$] In units of  $10^{-6}$~photons~s$^{-1}$~cm$^{-2}$.}
\endlastfoot
\end{longtable}

Next, we tried to derive the excess emission from the simultaneous fitting of HESS J1702--420 and background~1+2 spectra. 
The model parameters of the background spectra were linked between the HESS J1702--420 and the background~1+2 data. 
Furthermore, we modeled a possible emission from HESS J1702--420 as a power-law with $\Gamma=2.1$, same as that obtained by \citet{aharonian2008}, and included it in the model function for the HESS J1702--420 data. 
The fit was reasonably good with $\chi^2/{\rm d.o.f.} = 465.6/359$. 
To evaluate the significance of the additional power-law, we also performed a fit without the power-law component, which resulted in a slightly larger value of $\chi^2/{\rm d.o.f.} = 473.0/360$. 
The change of $\chi^2$ corresponds to F-value of $F^1_{360} =5.70$. 
A chance probability to obtain this F-value is 2.5\%, which means the significance of the additional power-law is marginal. 
However, it becomes insignificant if we consider the systematic error of the NXB and GRXE (5.9~\% and 20~\%, respectively, and $\simeq$~7.1~\% in total at 90~\% confidence range). 
The intrinsic X-ray flux in the 2--10~keV band is $3.4~^{+2.9}_{-2.7}{\rm (statistical)}~^{+2.0}_{-2.1}{\rm (systematic)} ~(10^{-13}~{\rm erg~s}^{-1}~{\rm cm}^{-2})$ with the 90~\% confidence range. 
Taking account both the statistical and systematic errors, we conclude that there is no significant excess emission in the entire XIS FOV ($7.4~\times~10^{-2}~{\rm deg}^{2}$) and the upper limit is $6.9~\times~10^{-13}~{\rm erg~s}^{-1}~{\rm cm}^{-2}$ with 90~\% confidence limit. 

\section{Discussion}
\label{sec:discuss}

\subsection{Comparison of X-ray and VHE gamma-ray fluxes}
\label{subsec:counterpart}

We detected two faint point sources and set an upper limit for diffuse emission. 
To consider the nature of the sources and the meaning of the upper limit, we evaluate a flux ratio ($F_{\rm TeV}/F_{\rm X}$), where $F_{\rm TeV}$ is the flux in the 1 -- 10~TeV and $F_{\rm X}$ in the 2 -- 10~keV band, respectively. 
If we assume that the VHE gamma-rays are produced via inverse Compton scattering of the cosmic microwave background (CMB) by electrons, and X-rays via synchrotron emission, the flux ratio corresponds to the ratio of the energy densities of CMB and the magnetic field. 
If the VHE gamma-rays are mostly produced by electrons, the ratio becomes smaller than $\sim 1$. 
On the other hand, if electrons are not involved, the ratio becomes larger than $\sim 1$. 
We list in Table \ref{tab:fluxratio} the flux ratios for a few selected PWN and SNR associated to the VHE gamma-ray sources and for dark particle accelerators. As seen in Table \ref{tab:fluxratio}, VHE gamma-ray associated PWN and SNR typically have a flux ratio less than 1 ($F_{\rm TeV}/F_{\rm X} < 1$). 

The two faint point sources detected in the XIS FOV are most likely not related to the VHE source. 
Their flux ratio, $F_{\rm TeV}/F_{\rm X}~\sim~10^3$, is too large to consider either of them as a X-ray counterpart of HESS~J1702--420. 
Furthermore, their X-ray fluxes are less than the upper limit of the diffuse X-ray flux. 
The diffuse emission fainter than the upper limit would be more plausible as the X-ray counterpart than two faint sources. 
Thus, these sources are not likely counterparts. 
However, it was difficult to deduce their nature from the Suzaku data due to the poor statistics and their origin is unknown. 

In the previous section, we derived the upper limit of the diffuse emission under the assumption that the source is uniformly extended in the XIS FOV. 
However, as shown in Figure \ref{fig:hessimage}, the VHE gamma-ray source is more extended than the XIS FOV. 
A possible X-ray counterpart may also be more extended than the XIS FOV. 
Thus we need to consider the source extension (outside the XIS FOV) to obtain the correct upper limit of the X-ray flux. 
We approximated that the rough size of the HESS source as a $0\fdg8 \times 0\fdg4$ rectangular (the blue box in Figure \ref{fig:hessimage}) and the X-ray surface brightness is constant over the entire rectangular region. 
Then, the upper limit for this source was estimated to be $2.7~\times~10^{-12}~{\rm erg~s}^{-1}~{\rm cm}^{-2}$ in the 2--10~keV band. 

If we use this upper limit, the flux ratio of this source becomes large ($F_{\rm TeV}/F_{\rm X}~>~12$). 
The (lower limit of the) ratio is comparable to those known as dark particle accelerators. 
We found that HESS J1702--420 is a dark particle accelerator. 

\subsection{Wide-band spectrum}
\label{subsec:wideband}

\begin{figure}
  \begin{center}
    \FigureFile(80mm,50mm){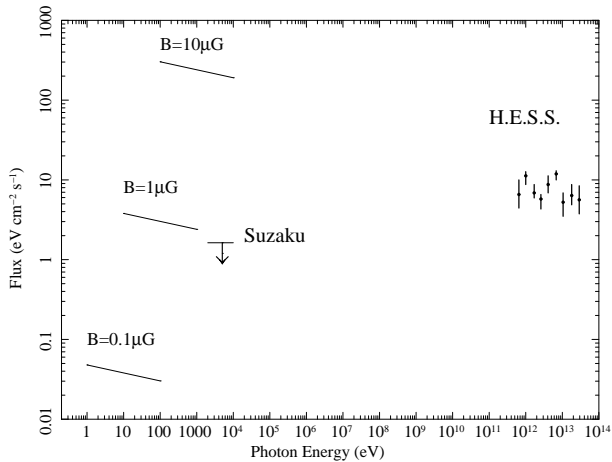}
  \end{center}
  \caption{Spectral energy distribution of HESS J1702--420 from the X-ray to VHE gamma-ray bands. The synchrotron X-ray emission expected for the simple one-zone model (using only CMB as target photon field for inverse Compton scattering) is indicated as a function of the magnetic field.}
  \label{fig:wideband}
\end{figure}

The wide-band spectral energy distribution of HESS~J1702--420 is shown in Figure \ref{fig:wideband}. 
The X-ray spectrum can be estimated from the VHE gamma-ray spectrum assuming a simple one-zone model and electron origin. 
In the most simple approach, the seed photon field is assumed to be only CMB, the Thomson cross section is taken for Compton scattering, and no cut-off in the electron spectrum is assumed (see \citet{balbo2010} for details of the model). 
The flux ratio corresponds to the ratio of the energy density of the seed photon and the local magnetic field. 
If we assume the photon index of the X-ray spectrum is same as that of the VHE gamma-ray spectrum ($\Gamma=2.1$), the X-ray flux is represented by a function of the local magnetic field. 
The estimated spectra for $B~=~0.1~\mu{\rm G}, 1~\mu{\rm G}, 10~\mu{\rm G}$ are overlaid on Figure \ref{fig:wideband}. 
According to the derived flux ratio, the local magnetic field would be $B < 0.9~\mu{\rm G}$, which is smaller than the typical value of $B \simeq 3 - 10~\mu{\rm G}$ on the Galactic plane. 
To resolve this discrepancy, several possibilities are conceivable: (1) a simple one-zone model does not hold for HESS~J1702--420, (2) the electron spectrum has a cut-off, and (3) the VHE emission is originated not from electrons but from protons. 

We will pursue the first possibility and will incorporate several modifications to the simple one-zone model. 
One is to consider seed photons other than CMB, i.e. optical/infrared photons from the interstellar radiation field. 
The contribution from the optical/infrared photons was estimated by \citet{porter2006} for the case of RX~J1713.7--3946 assuming two distances, 1~kpc and 6~kpc. 
We refer to the results for 6 kpc because it means relatively dense optical/infrared radiation field and is suited to estimate its maximum contribution.  
The distance is also compatible to those of many possible counterparts of the HESS sources located in the inner Galaxy (\cite{aharonian2006}). 
Interstellar optical and near-infrared emission may be estimated from the surface brightness of $3.5~\mu$m emission, which is almost the same between the regions of RX~J1713.7--3946 ($l=347\fdg33, b=-0\fdg47$) and HESS~J1702--420 (\cite{revnivtsev2006}). 
This means that the number density of the optical and near-infrared photons is also similar for these two sources. 
Here, we note that, according to \citet{porter2006}, the contribution of the optical and near-infrared field to the VHE emission is one tenth to that of the CMB in the case of RX~J1713.7--3946. 
Thus, we ignore the contribution from the optical and near-infrared emission around HESS~J1702--420 in the subsequent evaluation. 
On the other hand, interstellar far-infrared emission may have a significant contribution to the VHE emission. 
It may be estimated from the surface brightness at $\simeq 100~\mu$m. 
According to the IRAS $100~\mu$m map \footnote{http://irsa.ipac.caltech.edu/index.html}, the surface brightness of HESS~J1702--420 field is twice as large as that of RX~J1713.7--3946 field. 
When we compare the flux of the Compton up-scattered emission, we need to use different number densities of electrons for CMB and far-infrared photons, because the energy of Comptonizing electrons are different. 
This is equivalent to consider the electron energy distribution. 
In the case of HESS~J1702--420, the power-law index of the VHE emission is $\Gamma = 2.1$, while that of RX~J1713.7--3946 is 1.8--2.2. 
Because the power-law (photon) index is almost the same between HESS~J1702--420 and RX~J1713.7--3946, the slope of electron energy distribution is considered to be also same for the two sources. 
Incorporating all these parameters and applying the equation (1) of \citet{porter2006}, the Compton up-scattered emission of HESS~J1702--420 is estimated to be twice larger for the far-infrared photons. 
Because the CMB and far-infrared emission have similar contribution to the VHE emission in the case of RX J1713.7-3946, the Compton up-scattered emission of the CMB contribute one third to the VHE emission in HESS J1702-420. 
This means that the estimation of the magnetic field becomes $\sqrt{3} \simeq 1.7$ times as large, i.e. an upper limit of $1.7~\mu$G. 
This is still smaller than the typical value of the interstellar magnetic field on the Galactic plane. 

Another modification is the reduction of the Compton cross section in the Klein-Nishina regime. 
In the case of the simple one-zone model, we adopted a Thomson cross section for the Compton scattering for simplicity. 
However, this may not be very accurate to estimate the VHE emission due to the Compton up-scattering of CMB photons. 
Because the typical energy of the Comptonizing electrons is $E_{\rm IC} \simeq 20$~TeV, the CMB photon becomes at most 260~keV in the rest frame of the electron. 
This is almost a half of the rest-mass energy of an electron, and the Klein-Nishina cross section needs to be used. 
If we assume a 260~keV photon, the Klein-Nishina cross section becomes $\sigma \simeq 0.56 \sigma_{\rm T}$, where $\sigma_{\rm T}$ is the Thomson's cross section. 
If we incorporate the reduction of the cross section, this results in the increase of the number densities of electrons. 
Thus, the Klein-Nishina cross section tends to reduce the upper limit of the magnetic field and will even increase the discrepancy between the magnetic field on the Galactic plane and the upper limit. 

A cut-off in the electron spectrum is another possibility to explain our observational result. 
Typical electron energy to radiate VHE gamma-rays through the inverse Compton scattering is $E_{\rm IC} \simeq 20$~TeV, whereas that to radiate X-rays through synchrotron mechanism is $E_{\rm syn} \simeq 70$~TeV. 
Therefore, a cut-off between 20--70~TeV would reduce only the X-ray luminosity. 
Such a cut-off may be produced by (1) difference of electron life ($\tau \simeq 30$~kyr for $E_{\rm IC}$ vs $\tau \simeq 9$~kyr for $E_{\rm syn}$, where the energy loss is assumed to be dominated by the inverse Compton scattering of the CMB), or (2) the limitation of electron acceleration between $E_{\rm IC}$ and $E_{\rm syn}$. 

Because we could not find an X-ray counterpart of HESS~J1702--420, the nature of this source remains unknown. 
However, the number of similar sources (i.e. dark particle accelerator) is increasing. 
Thus, it might be a rather common source type in our Galaxy. 
\citet{yamazaki2006} suggest that an old SNR ($t~\simeq~3~\times~10^5$~yr) tends to have high flux ratio sometimes reaching $F_{\rm TeV}/F_{\rm X}~\sim~10^2$. 
Such an old SNR may be consistent with our results, because it preferentially emit soft X-rays which are easily absorbed by the interstellar matter on the Galactic plane. 
In this case, no significant X-ray emission may be observed. 
Thus an old SNR scenario may be one of the plausible possibilities consistent with the current X-ray and gamma-ray observations. 
A similar possibility is reported for HESS~J1745--303 by \citet{bamba2009}. 

\begin{longtable}{*{6}{c}}
\caption{Flux ratio of VHE gamma-ray objects}
\label{tab:fluxratio}
\hline
\hline
Name & $F_{\rm {TeV}}$\footnotemark[$*$] & $F_{\rm {X}}$\footnotemark[$\dagger$] & $F_{\rm {TeV}}/F_{\rm {X}}$ & Type & References \\
\hline
\multicolumn{1}{l}{Crab nebula} & 5.6 & $2.1 \times 10^5$ & $2.7 \times 10^{-3}$ & PWN & [1] [2] \\
\multicolumn{1}{l}{G0.9+0.1} & 0.2 & 58 & 0.3 & PWN & [3] \\
\hline
\multicolumn{1}{l}{RX J1713--3946} & 3.5 & 5400 & 0.06 & SNR & [4] [5] \\
\multicolumn{1}{l}{HESS J1813--178} & 0.9 & 70 & 1.3 & SNR & [6] [7] \\
\hline
\multicolumn{1}{l}{HESS J1616--508} & $1.7$ &  $<3.1$ & $>55$ & dark &  [6] [8] \\
\multicolumn{1}{l}{HESS J1745--303} & $0.52$  & $<2.1$ & $>25$ & dark (SNR?) &  [6] [9] \\
\multicolumn{1}{l}{HESS J1804--216} & $1.0$ & $<8.0$ & $>13$ & dark &  [6] [10] \\
\hline
\multicolumn{1}{l}{HESS J1702--420} & $3.1$ & $<27$ & $>12$ & dark & [11] this work \\
\hline
\multicolumn{6}{l}{\footnotemark[$*$] In units of $10^{-11}$~erg~s$^{-1}$~cm$^{-2}$ in the 1--10 TeV band.} \\
\multicolumn{6}{l}{\footnotemark[$\dagger$] In units of $10^{-13}$~erg~s$^{-1}$~cm$^{-2}$ in the 2--10 keV band.} \\
\multicolumn{6}{l}{[1] \citet{aharonian2004a}, [2] \citet{willingale2001}, [3] \citet{aharonian2005b} }\\
\multicolumn{6}{l}{[4] \citet{aharonian2004b}, [5] \citet{slane2001}, [6] \citet{aharonian2006} }\\
\multicolumn{6}{l}{[7] \citet{brogan2005}, [8] \citet{matsumoto2007}, [9] \citet{bamba2009}, }\\
\multicolumn{6}{l}{[10] \citet{bamba2007}, [11] \citet{aharonian2008}.}
\endlastfoot
\end{longtable}

\

We thank the referee for useful comments and suggestions. 
We also thank Keiko~Matsuta, Shigeo~Yamauchi, Kumiko~Morihana and Takao~Nakagawa for their useful information. 
We are grateful to Suzaku team and H.E.S.S. team. 
T. F. acknowledges the financial support from the Global Center of Excellence Program by the Japanese Ministry of Education, Culture, Sports, Science and Technology (MEXT) through the ``Nanoscience and Quantum Physics'' project of the Tokyo Institute of Technology, and Grant-in-Aid for JSPS Fellows by MEXT, no. $23\cdot9676$.


\begin{thebibliography}{}
  \bibitem[Abdo et al.(2010)]{abdo2010}
  Abdo, A. A., et al. 2010, \apjs, 188, 405 
  \bibitem[Aharonian \& Atoyan (1998)]{aharonian1998}
  Aharonian, F., \& Atoyan, A. M. 1998, ``Neutron Stars and Pulsars: Thirty Years after the Discovery'', 439 (arXiv:astro-ph/980309) 
  \bibitem[Aharonian et al.(2004a)]{aharonian2004a}
  Aharonian, F., et al. 2004a, \apj, 614, 897
  \bibitem[Aharonian et al.(2004b)]{aharonian2004b}
  Aharonian, F., et al. 2004b, \nat, 432, 75
  \bibitem[Aharonian et al.(2005a)]{aharonian2005a}
  Aharonian, F., et al. 2005a, Science, 307, 1938
  \bibitem[Aharonian et al.(2005b)]{aharonian2005b}
  Aharonian, F., et al. 2005b, \aap, 432, L25
  \bibitem[Aharonian et al.(2006)]{aharonian2006}
  Aharonian, F., et al. 2006, \apj, 636, 777
  \bibitem[Aharonian et al.(2008)]{aharonian2008}
  Aharonian, F., et al. 2008, \aap, 477, 353
  \bibitem[Balbo et al.(2010)]{balbo2010}
  Balbo, M., Saouter, P., Walter, R., Pavan, L., Tramacere, A., Pohl, M., \& Zurita-Heras, J. -A. 2010, \aap, 520, A111 
  \bibitem[Bamba et al.(2000)]{bamba2000}
  Bamba, A., Koyama, K., \& Tomida, H. 2000, \pasj, 52, 1157
  \bibitem[Bamba et al.(2001)]{bamba2001}
  Bamba, A., Ueno, M., Koyama, K., \& Yamauchi, S., 2001, \pasj, 53, L21
  \bibitem[Bamba et al.(2007)]{bamba2007}
  Bamba, A., et al. 2007, \pasj, 59, S209
  \bibitem[Bamba et al.(2009)]{bamba2009}
  Bamba, A., et al. 2009, \apj, 691, 1854
  \bibitem[Bell (1978)]{bell1978}
  Bell, A. R. 1978, \mnras, 182, 443
  \bibitem[Blandford, Ostriker (1978)]{blandford1978}
  Blandford, R. D., \& Ostriker, J. P. 1978, \apj, 221, L29
  \bibitem[Brogan et al.(2005)]{brogan2005}
  Brogan, C. L., Gaensler, B. M., Gelfand, J. D., Lazendic, J. S., Lazio, T. J. W., Kassim, N. E., \& McClure-Griffiths, N. M., 2005, \apj, 629, L105
  \bibitem[Chang et al.(2008)]{chang2008}
  Chang, C., Konopelko, A., \& Cui, W. 2008, \apj, 682, 1177
  \bibitem[Dubner et al.(1993)]{dubner1993}
  Dubner, G. M., Moffett, D. A., Goss, W. M., \& Winkler, P. F. 1993, \aj, 105, 2251
 \bibitem[Gallant (2007)]{gallant2007}
  Gallant, Y. A. 2007, \apss, 309, 197 
 \bibitem[Green et al.(1999)]{green1999}
  Green, A. J., Cram, L. E., Large, M. I., \& Ye, T. 1999, \apjs, 122, 207
  \bibitem[Green (2009)]{green2009}
  Green, D. A. 2009, Bull. Astr. Soc. India, 37, 45
  \bibitem[Guseinov et al.(2004)]{guseinov2004}
  Guseinov, O. H., Yerli, S. K., Ozkan, S., Sezer, A., \& Tagiyeva, S. O. 2004, Astronomical and Astrophysical Transactions, 23, 357 
  \bibitem[Ishisaki et al.(2007)]{ishisaki2007}
  Ishisaki, Y., et al. 2007, \pasj, 59, S113
  \bibitem[Kalberla et al.(2005)]{kalberla2005}
  Kalberla, M., et al. 2005, \aap, 440, 775
  \bibitem[Kokubun et al.(2007)]{kokubun2007}
  Kokubun, M., et al. 2007, \pasj, 59, S53
  \bibitem[Koyama et al.(1989)]{koyama1989}
  Koyama, K., Awaki, H., Kunieda, H., Takano, S., \& Tawara, Y. 1989, \nat, 339, 603
  \bibitem[Koyama et al.(1995)]{koyama1995}
  Koyama, K., Petre, R., Gotthelf, E. V., Hwang, U., Matsuura, M., Ozaki M., \& Holt, S. S. 1995, \nat, 378, 255
  \bibitem[Koyama et al.(1997)]{koyama1997}
  Koyama, K., Kinugasa, K., Matsuzaki, K., Nishiuchi, M., Sugizaki, M., Torii, K., Yamauchi, S., \& Aschenbach, B. 1997, \pasj, 49, L7
  \bibitem[Koyama et al.(2007)]{koyama2007}
  Koyama, K., et al. 2007, \pasj, 59, S23
  \bibitem[Matsumoto et al.(2007)]{matsumoto2007}
  Matsumoto, H., et al. 2007, \pasj, 59, S199
  \bibitem[Mitsuda et al.(2007)]{mitsuda2007}
  Mitsuda, K., et al. 2007, \pasj, 59, S1
  \bibitem[Nakajima et al.(2008)]{nakajima2008}
  Nakajima, H., et al. 2008, \pasj, 60, S1
  \bibitem[Porter et al.(2006)]{porter2006}
  Porter, T. A., Moskalenko, I. V., \& Strong, A. W. 2006, \apj, 648, L29
  \bibitem[Revnivtsev et al.(2006)]{revnivtsev2006}
  Revnivtsev, M., Sazonov, S., Gilfanov, M., Churazov, E., \& Sunyaev, R. 2006, \aap, 452, 169
  \bibitem[Serlemitsos et al.(2007)]{serlemitsos2007}
  Serlemitsos, P., et al. 2007, \pasj, 59, S9
  \bibitem[Slane et al.(2001)]{slane2001}
  Slane, P., Hughes, J. P., Edgar, R. J., Plucinsky, P. P., Miyata, E., Tsunemi, H., \& Aschenbach, B. 2001, \apj, 548, 814
  \bibitem[Takahashi et al.(2007)]{takahashi2007}
  Takahashi, T., et al. 2007, \pasj, 59, S35
  \bibitem[Tawa et al.(2008)]{tawa2008}
  Tawa, N., et al. 2008, \pasj, 60, S11
  \bibitem[Ubertini et al.(2005)]{ubertini2005}
  Ubertini, P., et al. 2005, \apj, 629, L109
  \bibitem[Uchiyama et al.(2008)]{uchiyama2008}
  Uchiyama, Y., et al. 2008, \pasj, 60, S35
  \bibitem[Willingale et al.(2001)]{willingale2001}
  Willingale, R.., Aschenbach, B., Griffiths, R. G., Sembay, S., Warwick, R. S., Becker, W., Abbey, A. F., \& Bonnet-Bidaud, J. M. 2001, \aap, 365, L212
  \bibitem[Yamauchi et al.(2009)]{yamauchi2009}
  Yamauchi, S., et al. 2009, \pasj, 61, S225
  \bibitem[Yamazaki et al.(2006)]{yamazaki2006}
  Yamazaki, R., Kohri, K., Bamba, A., Yoshida, T., Tsuribe, T., \& Takahara, F. 2006, \mnras, 371, 1975
  \bibitem[Yuasa et al.(2008)]{yuasa2008}
  Yuasa, T., et al. 2008, \pasj, 60, S207

\end{thebibliography}
\end{document}